\documentclass{PoS}

\title{RHIP, a Radio-controlled High-Voltage Insulated Picoammeter and its usage in studying ion backflow in MPGD-based photon detectors}

\ShortTitle{RHIP}

\author{M. Bari, B.~Gobbo, \speaker{S.~Dalla~Torre}\thanks{corresponding author}, M. Gregori, S.~Levorato, G.~Menon, F.~Tessarotto
\\INFN, Sezione di Trieste, Trieste, Italy
\\E-mail: \email{Silvia.DallaTorre@ts.infn.it}
}

\abstract{A picoammeter system has been developed 
and engineering. 
It consists in a current-voltage converter, based on an 
operational amplifier with very low input current, a high 
precision ADC, a radio controlled data acquisition unit 
and the computer-based control, visualization and 
storage. The 
precision is of the order of a tenth of picoampers and 
it can measure currents between electrodes at potentials up to 8 kV.  
The system is battery powered and a number of strategies have 
been implemented to limit the power consumption.  The 
system is designed for multichannel applications, up to 256 
parallel channels. 
The overall implementation is cost-effective to make the availability of multichannel setups easily affordable.
\par
The design, implementation and performance of the 
picoammeter system are described in detail as well as a an application:
the measurement of ion 
backflow in MPGD-based photon detectors.
}

\FullConference{5th International Conference on Micro-Pattern Gas Detectors (MPGD2017)\\
		22-26 May, 2017\\
		Philadelphia, USA}
\begin{document}
\section{Requirements}
The project arose from the need of very low current
measurements, with a precision of the order of a tenth of
picoampere, between points laying at potentials of the 
order
of a ten of kilovolts. Due to the high potentials involved, a
good insulation between case and circuitry is needed, as well as 
an efficient insulation from noise. An
automatic data read out and storage makes possible long 
measurement periods (hours or even days). The insulation
requirements impose battery supplies and low power consumption for long measuring periods. 
A reasonably low cost is required in order to make multichannel
systems easily affordable.
\par
The above listed requirements implied not trivial choices during the
development and prototyping of the apparatus.
\section{The device and its implementation}
\begin{figure}
  \begin{minipage}[b]{0.4\textwidth}
  \includegraphics[width=\textwidth]{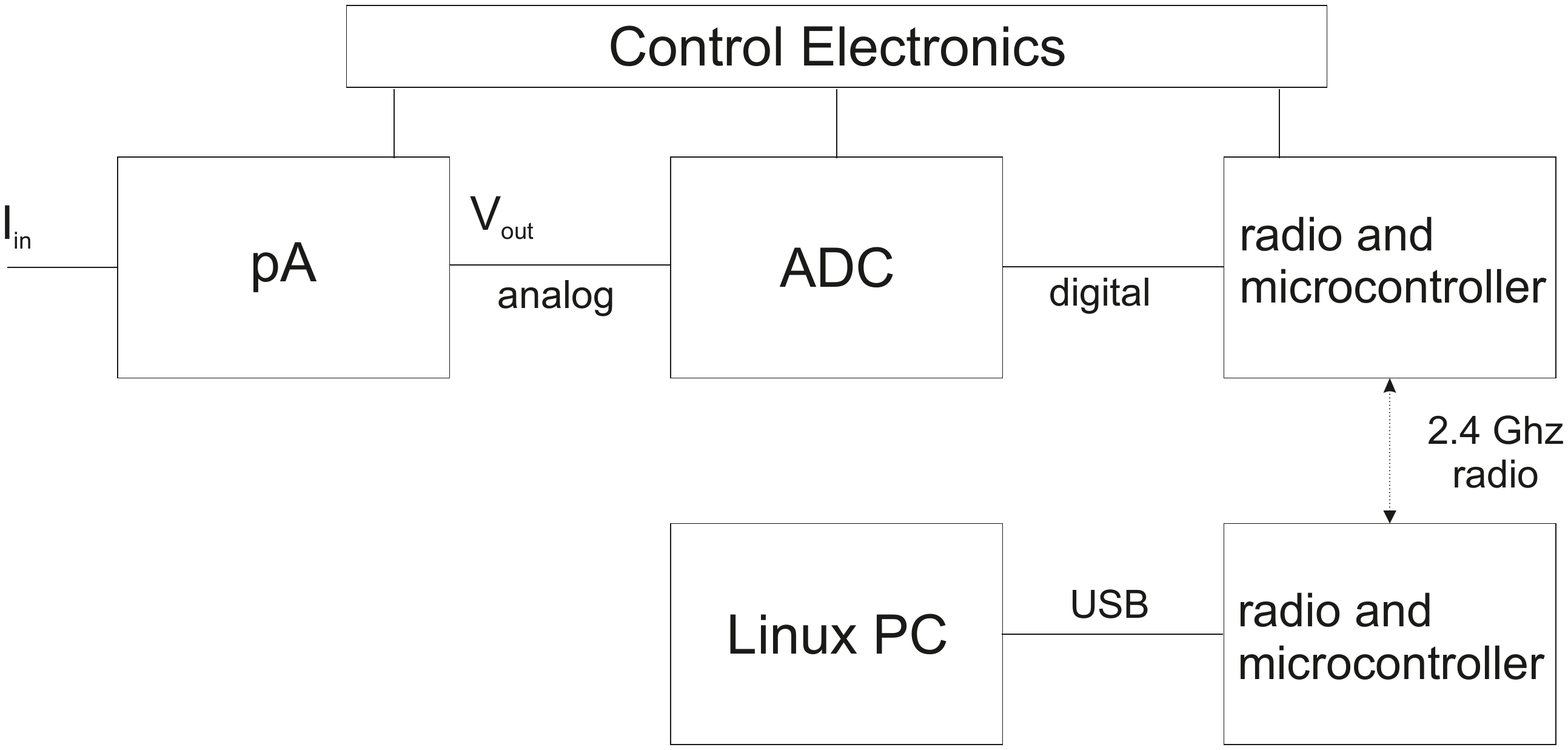}
    \caption{\label{fig:block-diagram}
    Block diagram of the picoammeter system.}
  \end{minipage}
  \hfill
\begin{minipage}[b]{0.4\textwidth}
\includegraphics[width=\textwidth]{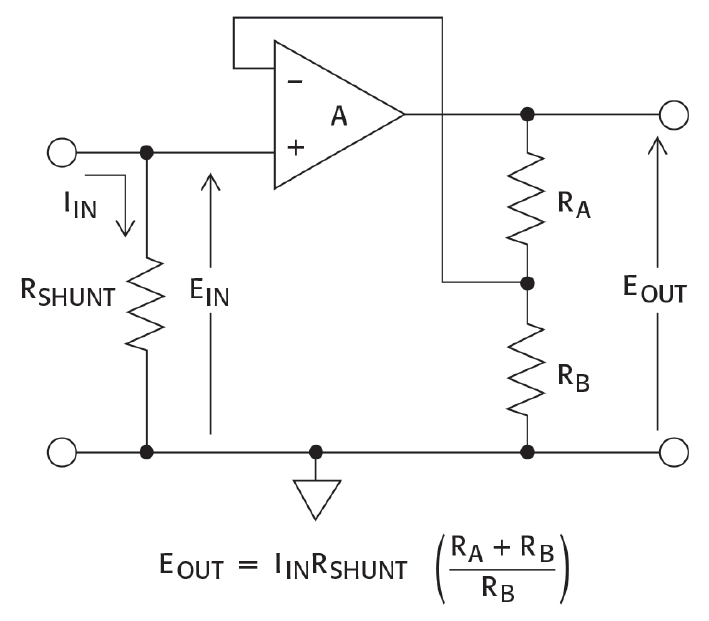}
    \caption{\label{fig:ampermeter} The ammeter scheme .}
\end{minipage}
\end{figure}

The picoammeter system can be split into five design
blocks (Fig.~\ref{fig:block-diagram}): a current-voltage converter, based on an operational
amplifier with very low input current; a high precision
ADC; a radio controlled data acquisition unit; the control
electronics; finally, the computer based overall control, visualization
and storage.
\par
The \textbf{current-voltage converter} scheme is a 
classical 
one based on an OPerational
Amplifier (OPA) with very low polarization current 
and feedback
resistor (Fig. ~\ref{fig:ampermeter}).
The selected OPA is a monolithic one with ultra-low 
input bias current, realized in JFET technology with
a typical input bias current of 40~fA, where the 
maximum value is 60~fA, at 25$^o$~C and supply voltage of $\pm$15~V.
Input offset voltage and input voltage drift are 
0.25~mV and 5~$\mu$volt/degree respectively.
The unity gain bandwidth is 1~MHz  and the  
slew rate is 3~V/$\mu$s.
Great attention was payed to guarantee
noise reduction, insulation of parts under high potential,
separation of the analogue part (basically the converter) from
the digital one (ADC, microcontroller, radio connection).
This part is powered with 
NiMH~PP3~8.2~V rechargeable batteries, compatible 
 with more than a week continuos usage.
\par
A \textbf{digitization stage} follows the current-voltage converter,
based on an analog to digital converter: a 24 bits (at least 20
bits of effective resolution) ADC is used in order to obtained
the required precision, equipped with SPI serial bus output.
The ADC chip, after a sample conversion, goes in standby
mode waiting for a serial readout request, a strategy
resulting in a sensitive reduction of the power consumption.
\par
A commercial microcontroller chip is used for the \textbf{radio control}: 
the selected one
can control serial buses as well as communicate via radio
with devices equipped with the same chip, in our application
a computer via its USB interface. The current absorption of
the chip is about 30 mA;  the chip must be kept always active,
because there is no option to wake it up from remote: it is 
the higher power consumption part
of the picoammeter system. A dedicated high capacity lithium
ions battery let this component work continuously for about a
week.
\par
The  \textbf{software to transmit and receive data} via the PC USB port 
was elaborated from an available example, whereas, \textbf{the software 
needed to interface the ADC via SPI} is entirely original and 
optimized for the picoammeter needs. 
\par
The \textbf{control electronics} consists of two parts: monitoring the 
status of the batteries and   switching
on and off the power of the device, performed without touching the case,
that lies at very high potential when connected.  The
battery status is controlled by low power comparators, that
control low energy LED diodes. LEDs light up when battery
voltages goes below a given threshold. Power is switched
on/off by latching relays, that absorb energy only during status
changes. The relay status change is induced by reed-relays
activated by the proximity of a magnet, so there is no
need of a contact with the picoammeter case \cite{konorov}. 
Finally, a low current LED indicates
the status of the device.
\par
\begin{figure}
  \begin{minipage}[b]{0.4\textwidth}
\includegraphics[width=\textwidth]{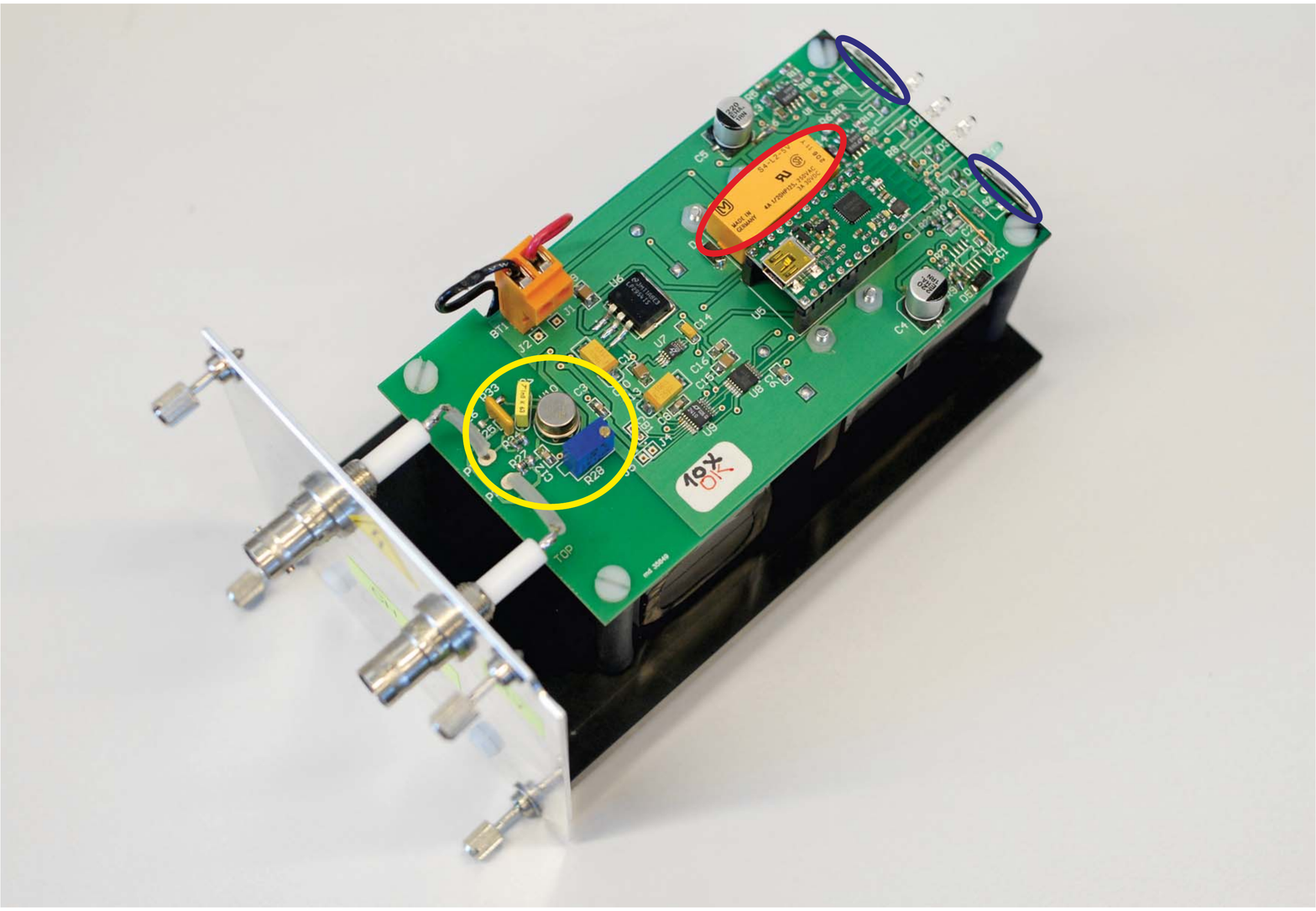}
    \caption{\label{fig:board}
    The picoammeter board. The analogic part is circled by yellow,
    latching relay in red circled, the two reed-relays in cyan.}
  \end{minipage}
  \hfill
\begin{minipage}[b]{0.4\textwidth}
\includegraphics[width=\textwidth]{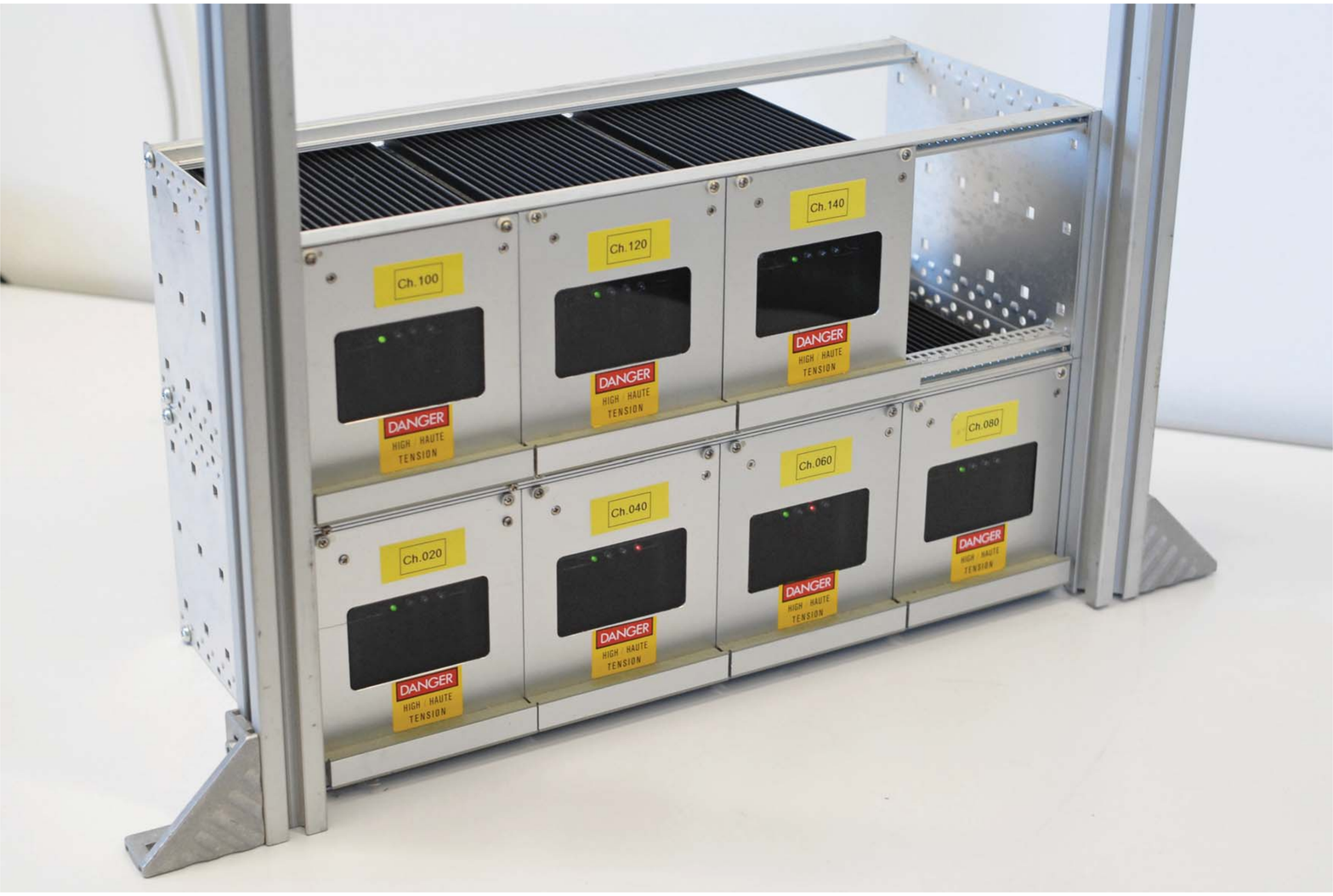}
    \caption{\label{fig:7} 
    Seven picoammeters  assembled in a mini crate.}
\end{minipage}
\end{figure}

A \textbf{set of seven picoammiter boards} has been assembled
and tested. One of the picoammiter boards 
is shown in Fig.~\ref{fig:board}:
some relevant details described in the previous 
section are in evidence. 
A set of seven picoammiter channels have been mechanically 
arranged in an euro
rack mini crate (Fig.~\ref{fig:7}).
\par
The picoampermeters have been operated with different feedback resistors and the output value has been averaged over different number of samples.
The typical measurement range is  $\pm$2~nA and they exhibit
precision of the order of 0.1 pA
with a good thermal stability.
\section{Data display and storage}
\begin{figure}
\begin{center}
\includegraphics[width=\textwidth]{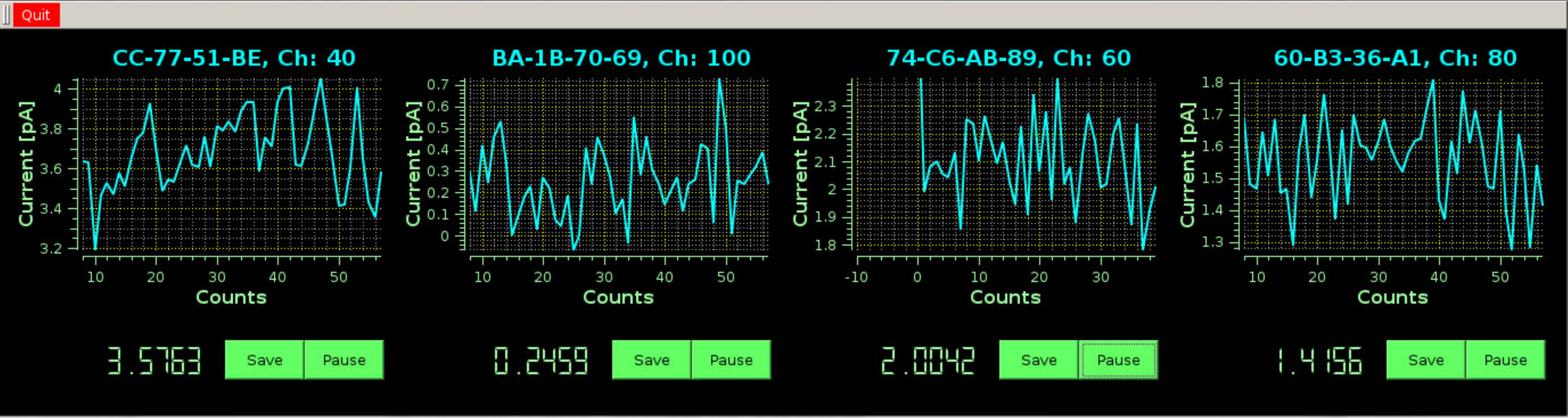}
\end{center}
\caption{\label{fig:gui}
The graphical interface, running on a Linux PC, connected to four
picoammeters. }
\end{figure}

Visualization (Fig. ~\ref{fig:gui}) and storage of read out data is
controlled by a program running on a Linux PC. The program
communicates as master, via radio link, to the picoammeter
ADC using one of its USB ports, acquiring
converted voltage measurements. The radio controller has 256
radiofrequency channels, so theoretically a single PC can
control up to 256 different picoammeters. The program is able
to identify the connected devices at start up; alternatively, it
is possible to specify the address of the devices to be contacted.
In addition, the program can store data from one or more
displayed channels to a text file, with the associated time
stamp. The storage to a file can be activate/deactivate at any
time. The program allows to define the time between two
consecutive measurements and the number of samples to be
averaged for a measurement; it also allows to set the timeout
to deactivate a not responding channel or to pause/resume a
picoammeter measurement.
\section{Measurement of the ion backflow rate in a hybrid MPGD by the RHIP system}
\begin{figure}
\begin{center}
\includegraphics[width=0.6
\textwidth]{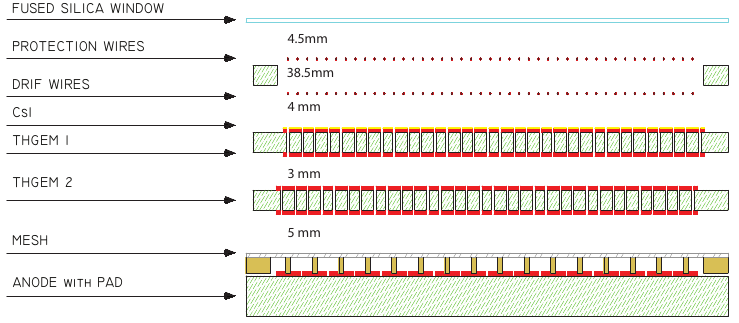}
\end{center}
\caption{\label{fig:hybrid}
Sketch of the hybrid single photon detector (image not to scale).}
\end{figure}

A novel single photon detector based on MPGD 
technologies~\cite{upgrade} has 
been developed for the upgrade
of the RICH-1 detector~\cite{rich1} of the 
COMPASS~\cite{compass} experiment at CERN SPS.
The new detector architecture consists in a hybrid
MPGD combination (Fig.~\ref{fig:hybrid}): two layers of THick GEMs
(THGEM)~\cite{thgem}, the first of which also acts as a reflective
photocathode (its top face is coated with a CsI film) are
coupled to a MicroMegas (MM)~\cite{mm} on a pad segmented
anode. The MM is resistive by an original implementation where 
the resistive MM is realized by discrete
elements: HV is applied to the anode pads , each one  protected 
by an individual resistor, while the signals are collected from a second 
set of pads, parallel to the first ones, embedded in the anode PCB 
where the signal is transfered by capacitively coupling. An 
important requirement is the limitation of the Ion BackFlow (IBF) 
rate to the CsI-coated photocathode: in fact, ion bombardment 
causes the reduction of the CsI quantum efficiency after an 
integrated charge of about 1~mC/cm$^2$~\cite{csi-qe-degradation}. 
The novel photon detector design takes into account this 
requirement and the architecture optimization has been obtained 
thanks to IBF measurement by the RHIP system. 
\par
\begin{figure}
  \begin{minipage}[b]{0.4\textwidth}
\includegraphics[width=\textwidth]{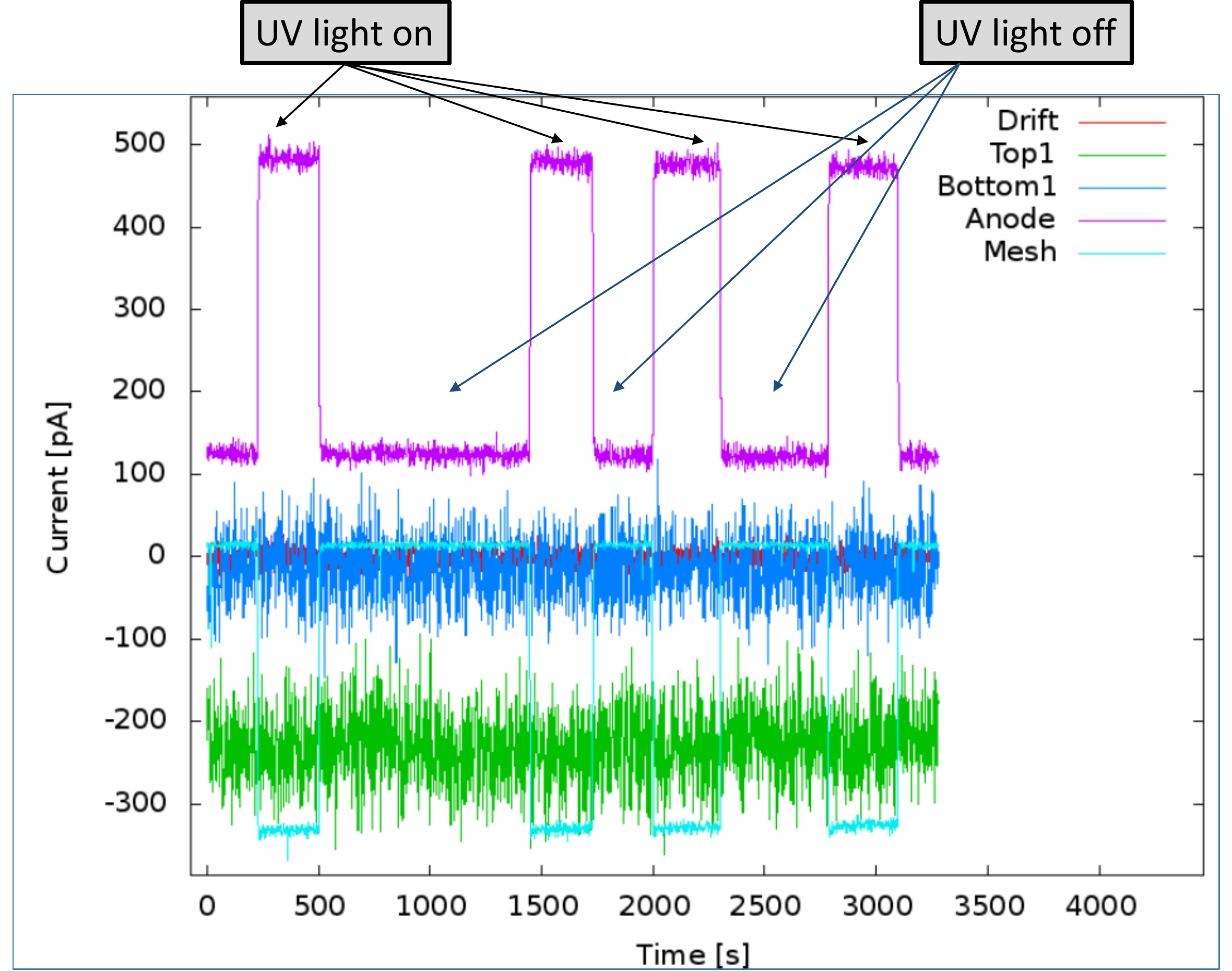}
    \caption{\label{fig:IBF} Currents on  hybrid detector electrodes versus 
    time; current off-sets make an easier visualization possible.
    }
  \end{minipage}
  \hfill
\begin{minipage}[b]{0.4\textwidth}
\includegraphics[width=\textwidth]{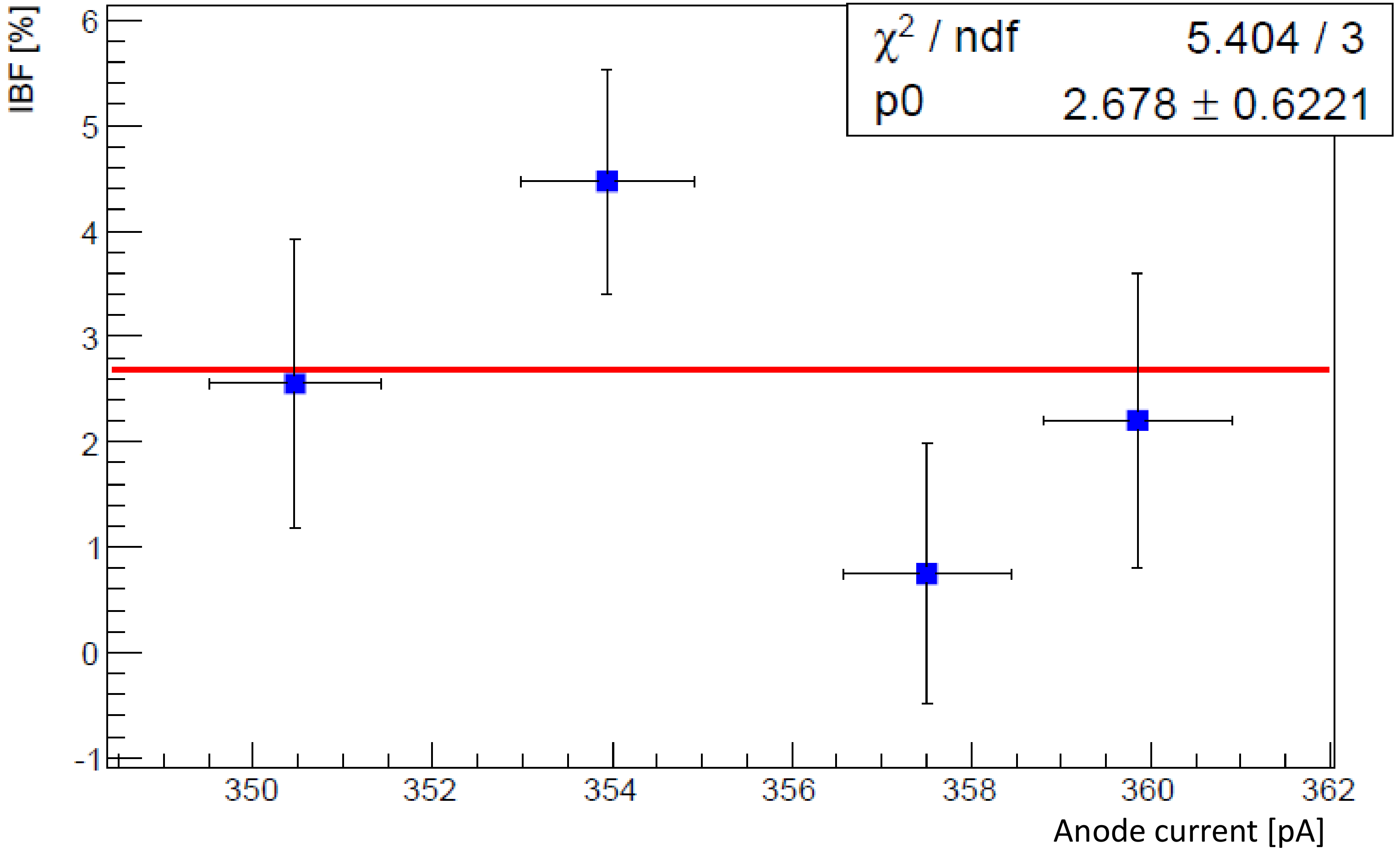}
\caption{\label{fig:IBF-results}
Measured IBF rate versus the anode current.}
\end{minipage}
\end{figure}

In a hybrid MPGD photon detector, IBF rate is measured by 
measuring the current of all the seven electrodes (drift wires, top 
and bottom faces of the two THGEMs, micromesh and anode pads) 
while illuminating the photocathode with UV light. A set of 
measurements performed alternating time intervals when light 
source is on to ones when it is off is shown in Fig.~\ref{fig:IBF}. The 
IBR rate  in the optimized configuration is presented in 
Fig.~\ref{fig:IBF-results} : IBR rate as low as 3\% is obtained.
\\
\par

\textbf{ACKNOWLEDGMENT}

The activity is partially supported by the H2020 project
AIDA2020 GA no. 654168.


\begin{thebibliography}{99}

\bibitem{konorov}
The use of reed-relays activated by magnets was
suggested by I.~Konorov, private communication.

\bibitem{upgrade}
F. Tessatotto et al., "The novel photon detectors based on MPGD technologies for the upgrade of COMPASS RICH-1", these proceedings and references therein.

\bibitem{rich1}
E. Albrecht et al., Nucl. Instr. and Meth. A 553 (2005) 215;
P. Abbon et al. , Nucl. Instr. and Meth. A 587 (2008) 371;
P. Abbon et al. , Nucl. Instr. and Meth. A 616 (2010) 21;
P. Abbon et al. , Nucl. Instr. and Meth. A 631 (2011) 26.

\bibitem{compass}
P. Abbon et al. , Nucl. Instr. and Meth. A 577 (2007) 455;
P. Abbon et al. ,  Nucl. Instr. and Meth. A 779 (2015) 69.

\bibitem{thgem}
L. Periale et al., Nucl. Instr. and Meth. A 478 (2002) 377; 
P. Jeanneret, PhD thesis, Neuchatel University, 2001; 
P.S. Barbeau et al, IEEE NS-50 (2003) 1285; 
R. Chechik et al, Nucl. Instr. and Meth. A 535 (2004) 303.

\bibitem{mm}
Y. Giomataris et al., Nucl. Instr. and Meth. A 376 (1996) 29.

\bibitem{csi-qe-degradation}
H. Hoedlmoser et al., Nucl. Instr. and Meth. A 574 (2007) 28.

\end{thebibliography}
\end{document}